\documentclass[aps,amsmath,amssymb,twocolumn,nofootinbib]{revtex4-1}

\pdfoutput=1

\usepackage[T1]{fontenc}

\usepackage{slashed}

\usepackage{amsmath,amssymb,url,bm}
\usepackage{amssymb,amsfonts}
\usepackage{rotating}
\usepackage{color}

\usepackage{amssymb}
\usepackage{amsmath}
\usepackage{graphicx,subfigure}
\usepackage{longtable}
\usepackage{verbatim}
\usepackage{hyperref}
\usepackage{bbold}
\usepackage{colortbl}

\usepackage{booktabs}

\usepackage{bm}

\usepackage{xcolor,colortbl}
\usepackage{array,multirow}

\definecolor{Gray}{gray}{0.85}
\definecolor{LightGreen}{rgb}{0.88, 1, 0.88}
\definecolor{Blue}{rgb}{0,1,1}
\definecolor{Lime}{rgb}{0,1,0}
\definecolor{LightCyan}{rgb}{0.88,1,1}
\definecolor{LightRed}{rgb}{1, 0.85, 0.85}
\definecolor{Red}{rgb}{1, 0, 0}
\definecolor{LightYellow}{rgb}{1, 1, 0.85}
\definecolor{Yellow}{rgb}{1,1,0.05}
\definecolor{LightBlue}{rgb}{0.87, 0.94, 1}
\definecolor{white}{gray}{1}
\definecolor{black}{gray}{0}

\definecolor{lightgray}{gray}{0.91}

\usepackage{array,mathtools,amssymb,booktabs}
\newcolumntype{C}{>{$}c<{$}}
\AtBeginDocument{
\heavyrulewidth=.16em
\lightrulewidth=.1em
\cmidrulewidth=.03em
\belowrulesep=.4ex
\belowbottomsep=0pt
\aboverulesep=.4ex
\abovetopsep=0pt
\cmidrulesep=\doublerulesep
\cmidrulekern=.5em
\defaultaddspace=.5em
}

\AtBeginDocument{
\heavyrulewidth=.16em
\lightrulewidth=.1em
\cmidrulewidth=.03em
\belowrulesep=.4ex
\belowbottomsep=0pt
\aboverulesep=.4ex
\abovetopsep=0pt
\cmidrulesep=\doublerulesep
\cmidrulekern=.5em
\defaultaddspace=.5em
}

\definecolor{Black}{gray}{0}
\definecolor{Gray}{gray}{0.85}
\definecolor{LightGray}{gray}{0.93}
\definecolor{LightGreen}{rgb}{0.88, 1, 0.88}
\definecolor{LightCyan}{rgb}{0.88,1,1}
\definecolor{LightRed}{rgb}{1, 0.25, 0.25}
\definecolor{LightRed}{rgb}{1, 0.85, 0.85}
\definecolor{LightYellow}{rgb}{1, 1, 0.85}
\definecolor{LightYellow}{rgb}{1, 1, 0.85}
\definecolor{LightBlue}{rgb}{0.87, 0.94, 1}
\definecolor{white}{gray}{1}

\newcolumntype{G}{>{\columncolor{LightGray}}c}

\renewcommand{\arraystretch}{1.25}
\newcommand*{\fp}[1]{{FP}${}_{#1}$}

\def\beq{\begin{equation}}
\def\eeq{\end{equation}}

\def\bea{\arraycolsep .1em \begin{eqnarray}}
\def\eea{\end{eqnarray}}
\def\Tr{{\rm Tr}}
\newcommand{\step}{\vspace{.5em}}

\def\eps{\epsilon}

\newcommand{\Nf}{N_f}

\def\eq#1{(\ref{#1})}

\def\s0#1#2{\mbox{\small{$ \frac{#1}{#2} $}}}
\def\0#1#2{\frac{#1}{#2}}

\def\grgl{\:\hbox to -0.2pt{\lower2.5pt\hbox{$\sim$}\hss}{\raise3pt\hbox{$>$}}\:}
\def\klgl{\:\hbox to -0.2pt{\lower2.5pt\hbox{$\sim$}\hss}{\raise3pt\hbox{$<$}}\:}

\newcommand{\cell}[2]{
  \begin{tabular}{#1} #2 \end{tabular}
}

\makeatletter
    \def\CT@@do@color{%
      \global\let\CT@do@color\relax
            \@tempdima\wd\z@
            \advance\@tempdima\@tempdimb
            \advance\@tempdima\@tempdimc
    \advance\@tempdimb\tabcolsep
    \advance\@tempdimc\tabcolsep
    \advance\@tempdima2\tabcolsep
            \kern-\@tempdimb
            \leaders\vrule
                    \hskip\@tempdima\@plus  1fill
            \kern-\@tempdimc
            \hskip-\wd\z@ \@plus -1fill }
    \makeatother

\allowdisplaybreaks[1]

\begin{document}

\begin{flushright}
\end{flushright}

\title{Asymptotic safety with Majorana fermions and   new large $N$ equivalences}

\author{Andrew D.~Bond}
\author{Daniel F.~Litim} 
\author{Tom Steudtner} 
\affiliation{Department of Physics and Astronomy, University of Sussex, Brighton BN1 9QH, UK}

\begin{abstract}
 Using Majorana fermions and elementary mesons we find  new massless  quantum field theories  with  weakly interacting ultraviolet fixed points.  We also   find  new classes of    large~$N$  equivalences  amongst    $SU$, $SO$ and $Sp$ gauge theories with different types of matter fields and Yukawa interactions. Results include a triality of  asymptotically  safe theories and dualities between asymptotically free matter-gauge theories with identical fixed points, phase diagrams,  and scaling exponents.  Implications for  conformal field theory and orbifold reductions   are indicated.
\end{abstract}

\maketitle

\section{\bf Introduction}

Equivalences or dualities between seemingly different  theories can provide valuable insights into the dynamics of quantum fields at weak and strong coupling. Well-known examples include  equivalences between   $SU$, $SO$, and $Sp$ gauge theories  in the limit where the rank  of the gauge group   is large \cite{Lovelace:1982hz},   electric-magnetic duality in supersymmetric theories \cite{Seiberg:1994pq}, the seminal AdS/CFT conjecture \cite{Maldacena:1997re},  or equivalences between theories related by orbifold/orientifold projections  \cite{Kachru:1998ys,Bershadsky:1998cb,Schmaltz:1998bg,Erlich:1998gb,Strassler:2001fs,Armoni:2003gp} where parent and child theories achieve coinciding perturbative expansions in the planar limit, and, under some conditions,   non-perturbative equivalence   \cite{Kovtun:2004bz,Dunne:2016nmc}.  Large $N$ equivalences have  also seen many applications in QCD-like theories including on the lattice \cite{Kovtun:2007py,Hanada:2011ju,Lucini:2012gg}.

On a different tack, the  discovery of interacting ultraviolet fixed points in QCD-like  theories,  first conjectured in   \cite{Bailin:1974bq},  
 has sparked a lot of interest 
\cite{Litim:2014uca,
Litim:2015iea,
Bond:2016dvk,
Codello:2016muj,
Bond:2017wut,
Bond:2017sem,
Buyukbese:2017ehm,
Bond:2017tbw,
Bond:2017lnq,
Bond:2017suy,
Kowalska:2017fzw,
Bond:2018oco,
Barducci:2018ysr,
Hiller:2019tvg,
Hiller:2019mou}.
It has led to a   general classification of $4d$ quantum field theories including necessary and sufficient  conditions and strict no-go theorems   for weakly  interacting fixed points   \cite{Bond:2016dvk,Bond:2018oco}. In the large $N$ limit,  proofs for asymptotic safety  with Dirac fermions are available  with   \cite{Bond:2017suy} and without supersymmetry \cite{Litim:2014uca,Bond:2017tbw,Buyukbese:2017ehm,Bond:2017lnq}. Key ingredients   are Yukawa interactions which can stabilise non-free gauge couplings \cite{Bond:2016dvk}. At finite $N$, these ideas are  used to UV complete the Standard Model \cite{Bond:2017wut,Kowalska:2017fzw,Barducci:2018ysr} and to study  aspects of flavour and  vacuum stability  \cite{Hiller:2019tvg,Hiller:2019mou}.

In this paper, we explain how asymptotic safety  materialises in theories with Majorana fermions and elementary mesons, and how this compares to settings with Dirac fermions.  With the help of  perturbation theory, the renormalisation group (RG),  negative dimensionality theorems, and ideas from string theory,  we   also put forward new classes of large $N$ equivalences  amongst  gauge-Yukawa theories with different gauge or  global symmetries, and  different types of matter fields.  Results include a triality of  asymptotically safe theories with $SU$, $SO$ or $Sp$ gauge groups  with identical phase diagrams and scaling exponents at  ultraviolet  critical points, and dualities between  asymptotically free gauge-matter  theories  with identical infrared  critical points, and more.

\begin{table}[b]
\renewcommand{\arraystretch}{1}
\begin{tabular}{cccc}
\toprule
\rowcolor{LightRed}
\bf Invariant & $\bm{SU(N)}$ & $\bm{SO(N)}$ & $\bm{Sp(N)}$ \\ 
\midrule
$d_R$ & $N$ & $N$ & $N$ \\
\rowcolor{LightGray} $C_2^R$ & $\frac12(N - 1/N)$ & $\frac14(N-1)$ & $\frac14(N+1)$ \\
$d_G$ & $N^2 -1$ & $\frac12 N (N - 1)$ & $\frac12 N (N + 1)$ \\
\rowcolor{LightGray}$C_2^G$ & $N$ & $\frac12({N-2})$ & $\frac12({N+2})$  \\ 
\toprule
\end{tabular}
	\caption{Dimensions and quadratic Casimirs of fundamental and adjoint representations	with 
	Dynkin index $S_2^R=\s012$.}
	\label{tab:reps}
\end{table}

\begin{table*}
\begin{center}
\begin{tabular}{ccccccccccc}
\toprule
\rowcolor{LightRed}
&$\ \bm{\alpha_g^*}\ $ 
&$\ \bm{\alpha_y^*}\ $ 
&$\bm{\alpha_u^*}$ 
&$\bm{\alpha_v^*}$ 
&
${}\ \ $
&$\ \bm{\vartheta_1}\ $ 
&$\ \bm{\vartheta_2}\ $ 
&$\ \bm{\vartheta_3}\ $ 
&$\ \bm{\vartheta_4}\ $
&\bf \ \ Type\ \  \\
\midrule
\bf \ \ FP${}_{\bm 1}\ \ $& 
$-\s0{8}{75}\eps$&0&0&0
&\cellcolor{white}
&$\ \ \0{16}{225}\, \eps^2$ 
&$\ \ \0{8}{25}\, \eps$ 
&$0$ 
&$0$ 
& \bf IR\\
\rowcolor{LightGray}
\bf FP${}_{\bm 2}$
&$\ \ \  \ \s0{52}{57}\eps\ \ $
&$\ \ \s08{19}\eps\ \ $
&$\ \ \s0{2\left(\sqrt{23}-1\right)}{19}\,\eps\ \ $
&$a_{{}_{\rm UV}}\,\eps$
&
&$\ \ -\0{104}{171}\,\eps^2\ \ $
&$\ \ \0{52}{19}\, \eps $
&$\ \ \0{8}{19} \sqrt{20 + 6 \sqrt{23}}\,\eps$
&$\ \ \ \ \0{16}{19}\sqrt{23}\,\eps\ \ $
&\bf UV \\
\bf FP${}_{\bm 3}$
&$-\s0{10}{3}\eps$
&$\ \ -\s0{4}{3}\eps\ \ $
&$\0{1-2\sqrt{3}}{3}\,\eps$
&$a_{{}{\rm IR}}\,\eps$
&\cellcolor{white}
&$\ \ \0{20}{9}\,\eps^2$
&$ -10\, \eps$
&$-8 \left(1 + \s043 \sqrt{3}\right)^{1/2}\,\eps$
&$-\0{32}{\sqrt{3}}\,\eps$ 
&\bf IR\\
\bottomrule
\end{tabular}
\end{center}
 \vskip-.5cm
 \caption{\label{FPs} Interacting fixed points and scaling exponents to leading order in $\eps$,
  with $a_{\rm UV}=\s0{2}{19}[(20+6\sqrt{23})^{1/2}-2\sqrt{23}]$ and $a_{\rm IR}=\s043\sqrt{3}-(1+\s043\sqrt{3})^{1/2}$. The fixed points FP${}_2$ (FP${}_{1,3}$) are UV (IR) and physical for
 small positive (negative) $\eps$, respectively.}
 \vskip-.3cm
\end{table*}

\section{\bf  Majorana fermions}

We consider non-abelian gauge theories coupled to Majorana fermions $\Psi_i$ and singlet complex scalar fields $H_{ij}$. Majorana fermions are their own charge conjugates $\Psi^c = \Psi$ whose left- and right-handed chiral components \begin{align}
\Psi = \frac{1}{\sqrt{2}}\left(\psi,\,\psi^c\right)^\intercal \label{majorana}
\end{align} 
relate to the same Weyl field $\psi$ with charge conjugation $\psi^c = \varepsilon \psi^*$ and  $\varepsilon=\left(\begin{smallmatrix} \phantom{+}0 & 1 \\ -1 & 0 \end{smallmatrix} \right)$. Real representations ensure that both Weyl components undergo identical gauge transformations, whose generators are purely imaginary and antisymmetric $t^a = - \left(t^a\right)^\intercal = - \left(t^a\right)^*$. For theories with chiral Yukawa interactions the requirement for real representations can be weakened to  include pseudo-real ones which are real up to a transformation $\left(t^a\right)^\intercal = - M\,t^a M^{-1}$. In either case  chiral gauge anomalies cancel due to the vanishing of 
\begin{equation} \label{eq:no-anomaly}
  d^{abc} \equiv \tfrac12 \Tr\left[t^a\left\{t^b,\,t^c\right\}\right] = 0\,. 
\end{equation}
To ensure strict perturbative control, we use a suitable large $N$ limit \cite{Veneziano:1979ec} which necessitates the Majorana fermions to be in the fundamental representation. The latter implies that  unitary or any of the exceptional gauge groups are excluded, which leaves us with   orthogonal or symplectic gauge groups. \step

{\bf Orthogonal gauge symmetry $\bm{SO(N)}$.} 
We begin with a theory of $\Nf$ Majorana fermions in the fundamental representation of an  $SO(N)$ gauge theory, interacting   with gauge-singlet complex scalar fields $H$. The theory has a global  $SU\left(\Nf\right)$ flavour symmetry with the Weyl components transforming in the fundamental and the scalars  in the two-index symmetric representation  $H_{ij} = H_{(ij)}$. The Lagrangian  reads
\beq\label{LM}
\begin{array}{rcl}
L&=&-\frac14 F^a_{\mu\nu}F^{a\mu\nu}
\\[1ex]&&
+\Tr\left({\psi^\dagger}\,  i\sigma^\mu D_\mu\, \psi \right)+ \Tr \left( \partial_\mu H^\dagger \partial^\mu H \right)
\\[1ex]&&
-\s012\, y \,\Tr\left(\psi^\intercal  h\, H \varepsilon\, \psi + {\psi^\dagger} h\,  H^\dagger \varepsilon\,\psi^*\right)
\\[1ex]&&
-u\,\Tr\,(H ^\dagger H )^2  -v\,(\Tr\,H ^\dagger H )^2
\end{array}
\eeq
where $F^a_{\mu\nu}$ denotes the non-abelian field strength, the trace sums over gauge and flavour indices, and gauge-contractions of fermion bilinears  $ \left(\chi\,\xi\right) = \chi_\alpha h^{\alpha\beta} \xi_\beta$ are symmetric with $h^{\alpha\beta} = h^{\beta \alpha}$ and  $h^{\alpha\beta}h_{\beta\gamma} =  \delta^\alpha_{\phantom{\alpha} \gamma}$.  The  four canonically marginal couplings  $\{g,y,u,v\}$ of the perturbatively renormalisable theory are   the gauge,  Yukawa,  single,  and  double trace quartic, respectively.

Next, we investigate the renormalisation group equations for the running couplings \cite{Machacek:1983tz,Machacek:1983fi,Machacek:1984zw,Luo:2002ti} and search for perturbative fixed points of the theory   \cite{Bond:2016dvk}.  Perturbative control  is achieved using a   Veneziano limit \cite{Veneziano:1979ec} where the dimension of the fundamental representation $d_R$ and the number of fermion flavours $\Nf$ are send to infinity while their ratio is kept fixed (see Tab.~\ref{tab:reps} for our conventions of group-theoretical parameters).  
The parameter
\beq\label{eps}
\eps=\frac{\Nf}{N}-\frac{11}{2}
\eeq
becomes continuous and may take any value within the range  $-\s0{11}{2}<\eps<\infty$. For $\eps<0$, the theory is asymptotically free, while asymptotic freedom is absent for $\eps>0$. Following   't Hooft  \cite{tHooft:1973alw},  we introduce rescaled   couplings suitable for a planar or large~$N$ limit
 \begin{equation}\label{thooft}
\alpha_x =
 \frac{d_R \, x^2}{(4 \pi)^2}, \qquad 
 \alpha_u =
 \frac{\Nf \,u}{(4 \pi)^2}, \qquad 
 \alpha_v = \frac{\Nf^2 \,v}{(4 \pi)^2}\,,
\end{equation}
where  $ x=g$ or $y$, and  beta functions  $\beta_i\equiv d\alpha_i/d\ln \mu$.  To the leading non-trivial orders in perturbation theory which is two loop in the gauge and one loop in the Yukawa and quartic beta functions, we find
\begin{equation}
\begin{array}{lcl}\label{rges:NcSO}
\beta_g&=& \alpha_g^2\left[\frac{2}{3} \eps\, + \left(\frac{25}{4} + \frac{13}{6}\eps\right)\alpha_g - \frac12 \left(\frac{11}{2} + \eps\right)^2 \alpha_y \right]\,,\\
\beta_y &=&  \alpha_y\left[\left(\frac{13}{2} + \eps \right)\alpha_y - 3 \alpha_g\right]\,,\\
\beta_u&=& 4 \alpha_u^2 + 2 \alpha_y \alpha_u - \left(\frac{11}{2} + \eps\right) \alpha_y^2\,,\\
\beta_v&=& 2 \alpha_v^2 + 8 \alpha_u \alpha_v + 6 \alpha_u^2 + 2 \alpha_y \alpha_v \,. 
\end{array}
\end{equation}
In any $4d$ quantum field theory, the weakly coupled fixed point solutions to $\beta_i=0$  are either of the Banks-Zaks or of the gauge-Yukawa type \cite{Bond:2016dvk,Bond:2018oco}.  For small $\eps$ they arise as a strict power series in $\eps$ where subleading terms up to order $\eps^n$  are obtained from the loop order $(n+1,n,n)$  in the gauge, Yukawa, and quartic beta functions  
\cite{Bond:2017lnq,Bond:2017tbw}. Also, any weakly coupled fixed point corresponds  to  a (unitary) conformal field theory \cite{Luty:2012ww}.

\begin{figure}[b]
  \includegraphics[width=.8\columnwidth]{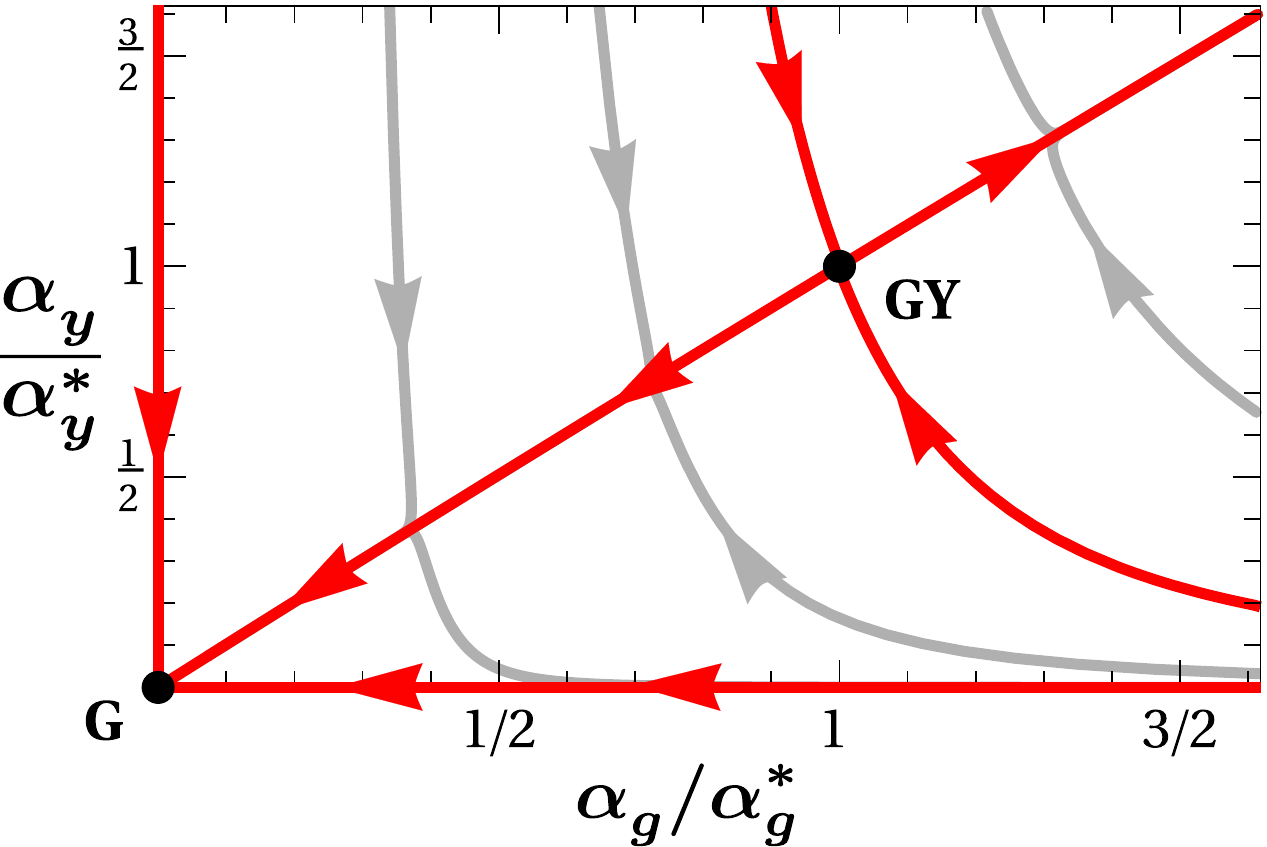}
  \caption{Phase diagram with asymptotic safety 
  ($\eps = 0.01$), projected onto the $(\alpha_g.\alpha_y)$ plane. Arrows point from the UV to the IR. Asymptotically safe trajectories emanate from the gauge-Yukawa fixed point \fp2 and run along a separatrix towards either a weakly or a strongly coupled IR regime. }
  \label{fig:AS-phase}
\end{figure}

Our results are summarised in Tab.~\ref{FPs}. In the regime with asymptotic freedom  the theory \eq{LM} with \eq{rges:NcSO} displays  a Banks-Zaks fixed point  \fp1. Infrared gauge-Yukawa fixed points are absent. In the regime where asymptotic freedom is lost, the gauge-Yukawa  fixed point \fp2 arises with $(\alpha_g^*,\alpha_y^*,\alpha_u^*,\alpha_u^*+\alpha_v^*) \approx   (0.91,0.42,0.40,0.13)\,\eps$ and a stable quantum vacuum \cite{Gildener:1975cj,Litim:2015iea}
\begin{equation}\label{majorana-stab}
\alpha^*_u \ge 0\,, \quad \alpha^*_u + \alpha^*_v \ge 0\,.
\end{equation}
A secondary fixed point  in the scalar sector does not lead to a stable vacuum and has been discarded. The universal  exponents $\vartheta_1<0<\vartheta_{2,3,4}$ establish that the fixed point is UV and the scaling power-law rather than logarithmic, and that the UV critical surface is one-dimensional corresponding to a single relevant coupling. 

The phase diagram with RG trajectories in the $(\alpha_g,\alpha_y)$-plane is displayed in Fig.~\ref{fig:AS-phase}. Switching on mass terms for the vector-like fermions or the scalars adds additional relevant directions (not shown), because perturbatively small anomalous dimensions cannot turn these into irrelevant operators. By the same token, higher dimensional interactions remain strictly irrelevant  \cite{Buyukbese:2017ehm}.

\begin{figure}[t]
  \includegraphics[width=.8\columnwidth]{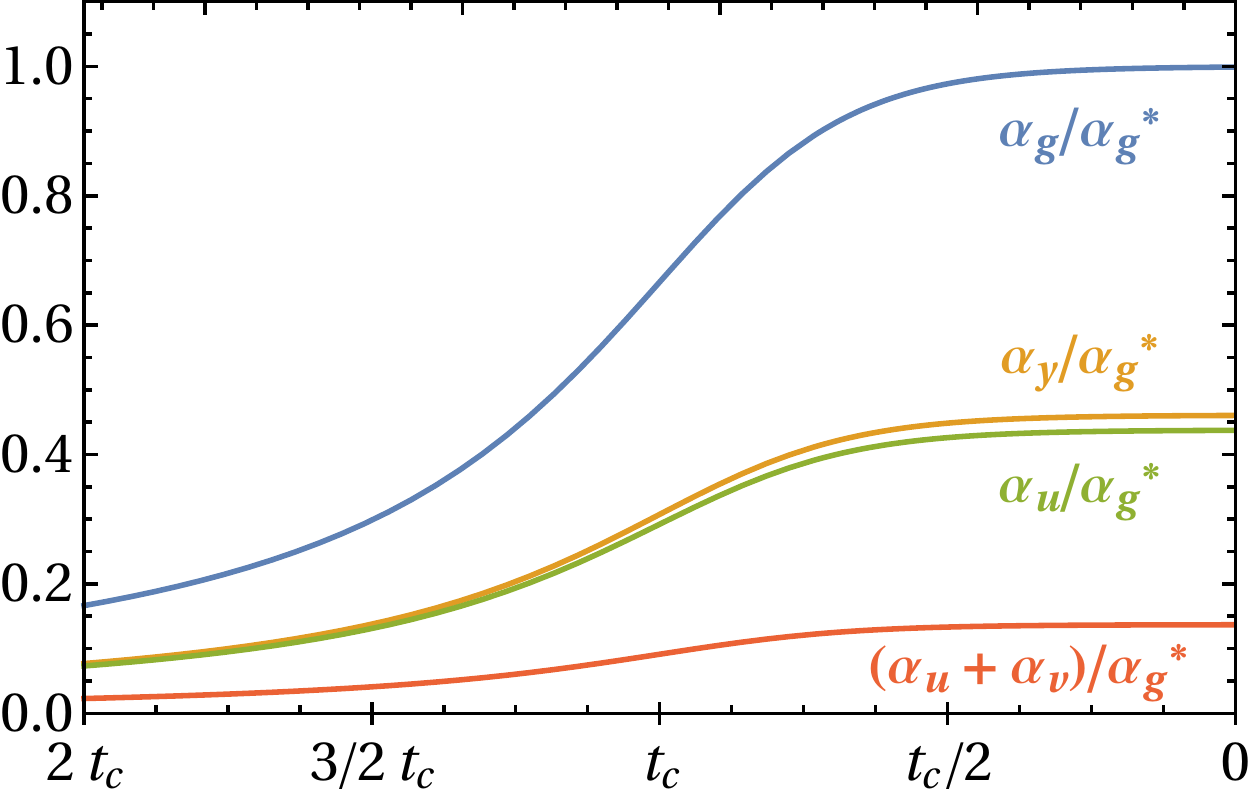}
  \caption{Cross-over of running couplings from asymptotic safety to infrared freedom in units of
  $\alpha_g^*$ with $ \mu_c=\Lambda\exp t_c$. }
  \label{fig:running}
\end{figure}

The separatrix  which connects the UV fixed point with the free IR fixed point is shown in Fig.~\ref{fig:running}. The   scale $\mu_c=\Lambda \exp t_c$  with    $\Lambda$ the high scale and $\alpha_g(t_c) = \tfrac23 \alpha_g^*$  \cite{Litim:2015iea} characterises the cross-over between the two fixed points and is the analogue of $\Lambda_{\rm QCD}$ in QCD. A second separatrix exists towards a regime with strong coupling and confinement in the IR (not shown). 

  Finally, we note that all previously known quantum field theories in four dimensions with exact asymptotic safety  involve unitary gauge symmetry and Dirac fermions   \cite{Litim:2014uca,Bond:2017tbw,Bond:2017suy,Bond:2017lnq}. In this light, the theory \eq{LM} with \eq{rges:NcSO}  offers the first proof of existence for  asymptotic safety in gauge theories with $SO(N)$ gauge symmetry, and in theories with Majorana fermions.  \step

{\bf Symplectic gauge symmetry $\bm{Sp(N)}$.} 
Next, we turn to  a theory of $\Nf$ Majorana fermions in the fundamental representation of an  $Sp(N)$ gauge theory  interacting with gauge-singlet complex scalar fields $H$.  In our conventions $N$ is an even integer, and $Sp(2)\simeq SO(3) \simeq SU(2)$. The theory has a global  $SU\left(\Nf\right)$ flavour symmetry with Weyl components transforming in the fundamental and $H$ 
in the two-index antisymmetric representation  $H_{ij} = H_{[ij]}$. To avoid a Witten anomaly \cite{Witten:1982fp} $\Nf$ has to be an even integer as well. The perturbatively renormalisable  Lagrangian of the theory takes the form
\beq\label{LM2}
\begin{array}{rcl}
L&=&-\frac14 F^a_{\mu\nu}F^{a\mu\nu}
\\[1ex]&&
+\Tr\left({\psi^\dagger}  i\sigma^\mu D_\mu\, \psi \right)+ \Tr \left( \partial_\mu H^\dagger \partial^\mu H \right)
\\[1ex]&&
-\s012\, y \,\Tr\left(\psi^\intercal f\, H \varepsilon\, \psi + {\psi^\dagger}f\, H^\dagger \varepsilon\,\psi^*\right)
\\[1ex]&&
-u\,\Tr\,(H ^\dagger H )^2  -v\,(\Tr\,H ^\dagger H )^2
\end{array}
\eeq
where we recall that gauge-contractions of fermion bilinears  $ \left(\chi\,\xi\right) = \chi_\alpha f^{\alpha\beta} \xi_\beta$ are antisymmetric with $f^{\alpha\beta} = - f^{\beta \alpha}$ and  $f^{\alpha\beta}f_{\beta\gamma} = - \delta^\alpha_{\phantom{\alpha} \gamma}$.

A  Veneziano limit  is established using \eqref{eps} and  rescaled couplings \eqref{thooft}.  Introducing the parameter $\eps$ as in \eq{eps}, we find the RG beta functions  for all couplings to the leading non-trivial order in perturbation theory.  Denoting the `t~Hooft couplings \eq{thooft} for the theories  \eq{LM} and  \eq{LM2} as $\alpha_{i}^{SO}$ and $\alpha_{i}^{Sp}$ respectively,  we find the remarkable result that  the RG beta functions  \eqref{rges:NcSO} for the theory  \eq{LM} agree exactly with those of the theory  \eq{LM2},  after the identification of couplings 
 \beq \label{id}\alpha_{i}^{SO} =  \alpha_{i}^{Sp}\,.
 \eeq
Consequently the fixed points and scaling exponents (\fp1 and \fp2 in Tab.~\ref{FPs}), and the RG trajectories and phase diagrams (Figs.~\ref{fig:AS-phase} and \ref{fig:running}) of the theories \eq{LM}  and  \eq{LM2} are {\it identical} in the Veneziano limit.  However, we also note that the equivalence is mildly violated  beyond the Veneziano limit  at large yet finite $N$ and $\Nf$ due to subleading corrections of order $1/N$ and $1/\Nf$ which arise with the same magnitude but opposite sign. 

Finally, we emphasise that the theory \eq{LM2} yields the first  rigorous example for  asymptotic safety in a symplectic gauge theory coupled to matter. The result thus establishes  that  asymptotic safety can be achieved in $4d$ quantum field theories with any of the non-exceptional gauge groups, and for sufficiently large $N$.

\section{\bf Dirac fermions}

Next, we consider theories of $\Nf$ Dirac fermions $\Psi_i$  interacting with non-abelian gauge fields and gauge-singlet complex scalar fields $H_{ij}$.  The theories have  a global $SU(\Nf)\times SU(\Nf)$ flavour symmetry with the elementary scalars $H$ transforming in the bifundamental.  The perturbatively renormalisable Lagrangian  is given by
 \beq\label{LD}
\begin{array}{rcl}
L&=&-\frac14 F^a_{\mu\nu}F_a^{\mu\nu}
\\[1ex]&&
+\Tr\left(\overline{\Psi}\,  i\slashed{D}\, \Psi \right)+ \Tr \left( \partial_\mu H^\dagger \partial^\mu H \right)
\\[1ex]&&
-y \,\Tr\left(\overline\Psi_L H \Psi_R + \overline{\Psi}_R H^\dagger \Psi_L\right)
\\[1ex]&&
-u\,\Tr\,(H ^\dagger H )^2  -v\,(\Tr\,H ^\dagger H )^2
\end{array}
\eeq
where $F^a_{\mu\nu}$ denotes the non-abelian field strength, the trace sums over all indices and the decomposition $\Psi=\Psi_L+\Psi_R$ with $\Psi_{L/R}=\frac 12(1\pm \gamma_5)\Psi$ is understood. Due to the fermions being vector-like, gauge-anomalies cancel by design and no restriction on their representations apply. In addition to the gauge coupling $g$ and the Yukawa coupling $y$, we  observe two independent quartic self interactions $u$ and $v$, which provides  us with a set of  four canonically marginal couplings $\{g,y,u,v\}$. In the following, we consider the Dirac fermions in the  fundamental gauge representation of $SU(N)$, $SO(2N)$ and $Sp(2N)$. We also establish a Veneziano limit using the  parameter \eqref{eps} and adopt the same set of `t~Hooft couplings \eqref{thooft}   as in the cases with  Majorana fermions.   \step

{\bf Unitary gauge symmetry $\bm{SU(N)}$.} 
For unitary gauge groups, the theory \eq{LD} has been studied  in a number of works \cite{Terao:2007jm,Kaplan:2009kr,Litim:2014uca,Litim:2015iea,Bond:2017tbw}. In the regime with asymptotic freedom, it can display a Banks-Zaks fixed point. Once  asymptotic freedom is lost, it develops a weakly interacting  asymptotically safe UV fixed point  \cite{Litim:2014uca} with a stable quantum vacuum \cite{Litim:2015iea}. The corresponding UV conformal window has been determined up to the complete next-to-next-to-leading order in perturbation theory which is three loop in the gauge and two loop in the Yukawa and quartic couplings  
\cite{Bond:2017tbw}. \step

The  main observation here is that the theory \eq{LD} with $SU$ gauge symmetry and Dirac fermions  is intimately related to the theories  \eq{LM} with $SO$ and to  \eq{LM2} with $Sp$ gauge symmetry and Majoranas. Introducing the parameter $\eps$ as in \eq{eps} and denoting the  couplings \eq{thooft} for the theory  \eq{LD}  with unitary gauge symmetry as  $\alpha_{i}^{SU}$, we find that   beta functions in the Veneziano limit are {\it identical} to those of the theories \eq{LM} and \eq{LM2},  given by \eqref{rges:NcSO}, provided we rescale the `t~Hooft couplings by a factor of two,
\begin{equation} \label{super-universality}
   \alpha_{i}^{SO}= \alpha_{i}^{Sp}=2\,\alpha_{i}^{SU}\,.
\end{equation} 
Consequently fixed points are either of the Banks-Zaks (\fp1) or the gauge-Yukawa-type (\fp2) and take the values given in Tab.~\ref{FPs} after rescaling. Phase diagrams and RG trajectories in these theories are also identical up to \eqref{super-universality}, and given by Figs.~\ref{fig:AS-phase} and ~\ref{fig:running}. Most notably, universal scaling exponents, which are insensitive to the normalisation of couplings, are identical between the two theories, and take the values given in Tab.~\ref{FPs}. \step

{\bf $\bm{SO(2N)}$ gauge symmetry.} 
 Next, we consider  settings with $\Nf$ Dirac fermions in the fundamental representation of $SO(2N)$ gauge symmetry, again coupled to scalars in the bifundamental two-index representation  of the global $SU(\Nf)\times SU(\Nf)$ flavour symmetry, and with action \eq{LD}. Notice that since Dirac fermions have twice as many degrees of freedom as Majorana fermions,  and to ensure that the  definition for the small parameter $\eps$ \eq{eps} remains unchanged, the dimension of the gauge group has been taken twice as large as in  the case with $SU$ gauge symmetry.  Then, to the leading order in perturbation theory and in the Veneziano limit, we find
\begin{equation}
\begin{array}{lcl}\label{rges:DcSO}
\beta_g&=& \alpha_g^2\left[\frac{2}{3} \eps\, + \left(\frac{25}{4} + \frac{13}{6}\eps\right)\alpha_g - \frac12 \left(\frac{11}{2} + \eps\right)^2 \alpha_y \right]\,,\\
\beta_y &=&  \alpha_y\left[\left(\frac{15}{2} + \eps \right)\alpha_y - 3 \alpha_g\right]\,,\\
\beta_u&=& 8 \alpha_u^2 + 4 \alpha_y \alpha_u - \left(\frac{11}{2} + \eps\right) \alpha_y^2\,,\\
\beta_v&=& 4 \alpha_v^2 + 16 \alpha_u \alpha_v + 12 \alpha_u^2 + 4 \alpha_y \alpha_v 
\end{array}
\end{equation}
for models \eq{LD} with orthogonal gauge symmetry. \step

In stark contrast to the previous examples, no interacting UV fixed points are found  as soon as asymptotic freedom is absent  \cite{Bond:2016dvk}.  However, the beta functions \eq{rges:DcSO} admit interacting fixed points provided the theory is asymptotically free ($\eps<0)$. These are either of the Banks-Zaks  (\fp1) or  of the gauge-Yukawa-type (\fp3), with fixed point coordinates and scaling exponents  summarised in Tab.~\ref{FPs}.  The  gauge-Yukawa  fixed point \fp3 at $(\alpha_g^*,\alpha_y^*,\alpha_u^*,\alpha_u^*+\alpha_v^*) \approx  -(3.33, 1.33, 1.23, 0.58)\,\eps$ also displays a stable quantum vacuum \eq{majorana-stab}. 

\begin{figure}
  \includegraphics[width=.75\columnwidth]{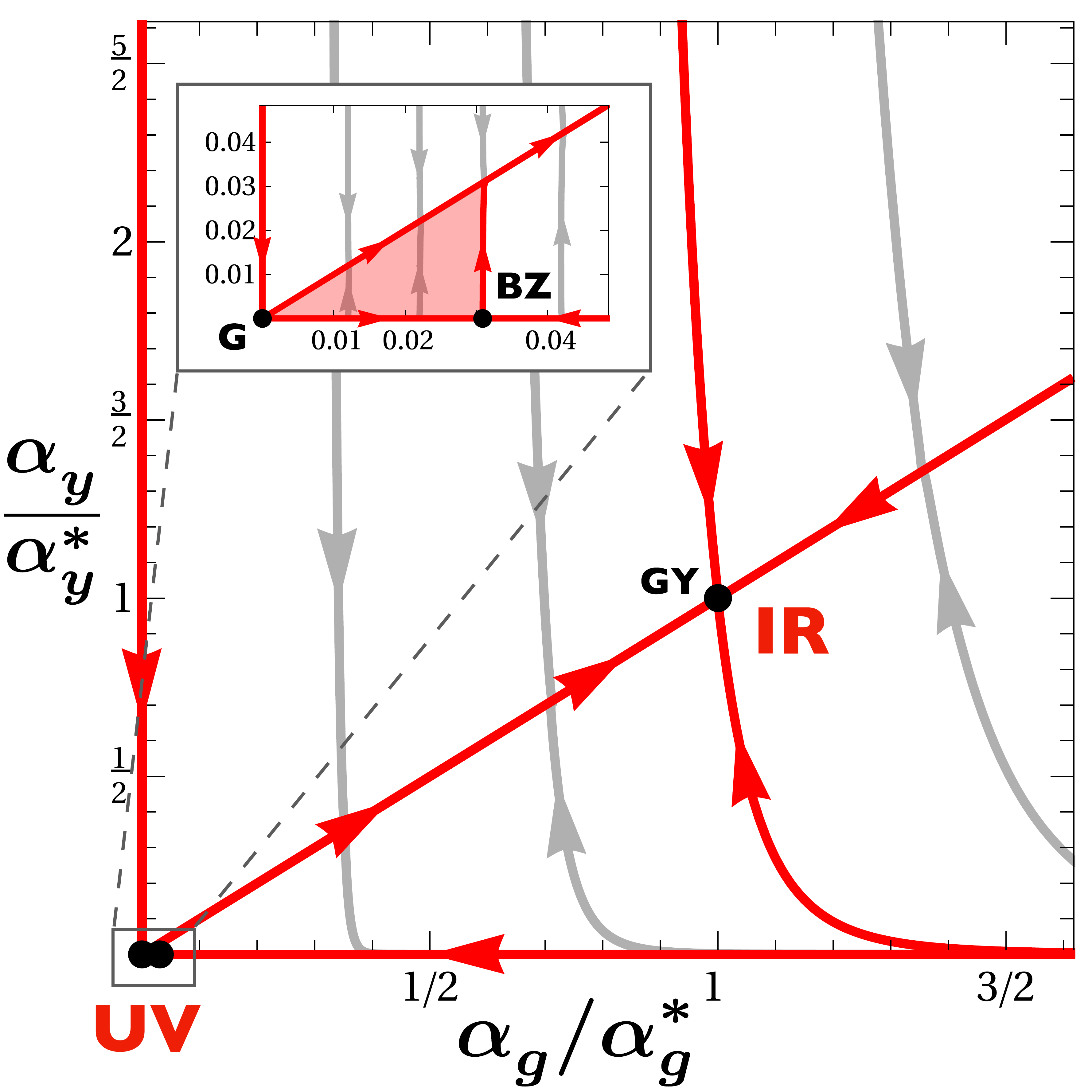}
\vskip-.3cm
  \caption{Phase diagram with asymptotic freedom ($\eps =-0.01$)
   projected onto the $(\alpha_g,\alpha_y)$ plane. Arrows point from the UV to the IR. Dots show  the Gaussian (G), the  Banks-Zaks (BZ), and the infrared gauge-Yukawa (GY)   fixed points. The inset highlights  the two-dimensionality of the UV critical surface  (red shaded area) which becomes effectively one-dimensional in the cross-over to the IR.}
  \label{fig:noAS-phase}
\end{figure}

The universal  exponents $0<\vartheta_{1,2,3,4}$  establish that the fixed point \fp3 is  fully attractive in all canonically marginal couplings  thus corresponding to an IR sink  \cite{Bond:2017wut}, and that the scaling is power-law rather than logarithmic.  The   phase diagram  in regimes with asymptotic freedom is shown in Fig.~\ref{fig:noAS-phase}.  We notice that the Banks-Zaks fixed point is parametrically small compared to the gauge-Yukawa fixed point (see the inset in Fig.~\ref{fig:noAS-phase}). It implies that the UV critical surface at the Gaussian fixed point, which is two-dimensional, {\it effectively} becomes 
one-dimensional, given by the separatrix connecting the Gaussian and the GY fixed point.\step

{\bf $\bm{Sp(2N)}$ gauge symmetry.} 
Finally, we turn to the models \eq{LD}  with $\Nf$ Dirac fermions in the fundamental of  $Sp(2N)$  gauge symmetry,  coupled to scalars in the bifundamental two-index representation  of the global $SU(\Nf)\times SU(\Nf)$ flavour symmetry. Using   \eq{eps},  introducing couplings as in \eq{thooft}, and following the same steps as before, we find once more  that the  beta functions in the Veneziano limit come out  {\it identical} to those found in  \eq{rges:DcSO}  after  a straight identification of couplings \eq{id}. It follows that the running of couplings, the phase diagrams, and the conformal critical points of theories \eq{LD} are identical,  irrespective of whether we impose  an orthogonal or symplectic gauge symmetry. 

\section{\bf Large $N$ equivalences}

In this section, we investigate the kinematical equivalences detected in the previous sections from the point of view of weak-coupling dualities and orbifold projections, and discuss implications for asymptotic safety.

{\bf Negative dimensionality theorems.} 
Some of our results can be understood with  the help of  so-called negative dimensionality theorems  
 \cite{King:1971rs,Mkrtchian:1981bb,Cvitanovic:1982bq,Cvitanovic:2008zz,Mkrtchyan:2010tt}. They state  that for any  $SO(L)$ invariant scalar there exists a corresponding $Sp(L)$ invariant scalar, and vice versa, obtained by exchanging symmetrisations and antisymmetrisations, replacing the $SO(L)$ symmetric bilinear invariant $h_{\alpha\beta}$ by the $Sp(L)$ antisymmetric bilinear invariant $f_{\alpha\beta}$, and replacing $L$ by $-L$. Similarly, for any $SU(M)$ invariant scalar exchanging symmetrisations and antisymmetrisations is equivalent to replacing $M$ by $-M$. Schematically, we write the theorems as
\begin{equation}\label{SoSp}
SO\left(L\right) = \overline{Sp}\left(-L\right), \qquad SU\left(M\right) = \overline{SU}\left(-M\right)\,,
\end{equation}
where overlines  indicate the transposition of  Young tableaux for all representations, corresponding precisely to the interchange of symmetrisation and antisymmetrisation \cite{King:1971rs,Mkrtchian:1981bb,Cvitanovic:1982bq,Cvitanovic:2008zz,Mkrtchyan:2010tt}.\step

 \begin{figure}[t]
  \includegraphics[width=.8\columnwidth]{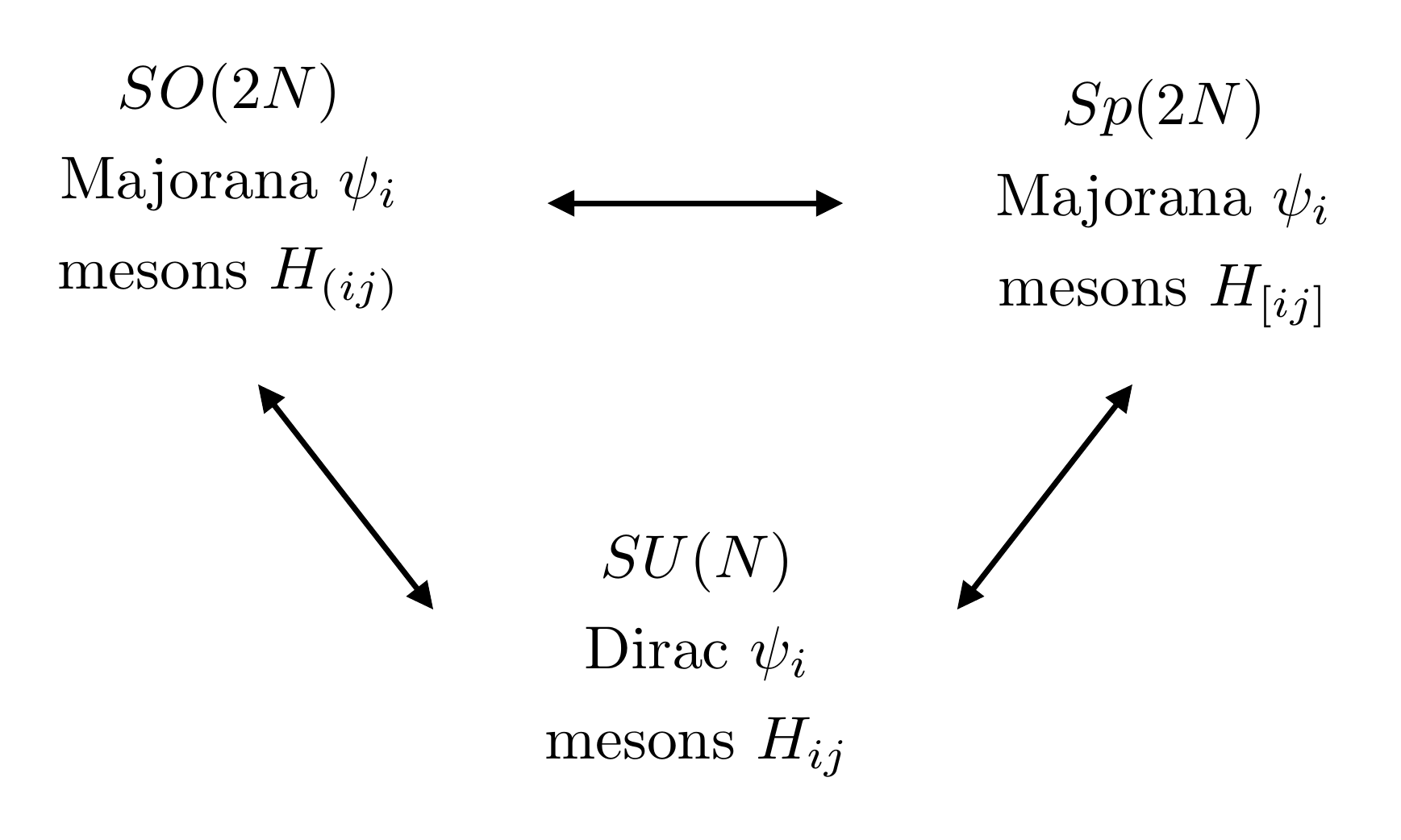}
 \vskip-.3cm
\caption{ Triality of asymptotic safety, and large~$N$ equivalences amongst  
 matter-gauge theories with  global $SU(2\Nf)$ flavour symmetry and Majorana fermions (top), and a theory with a global $SU(\Nf)\times SU(\Nf)$ flavour symmetry and Dirac fermions (bottom). 
 The horizontal arrow  emphasises  that RG flows, phase diagrams, and  critical points are identical. Top-down arrows indicate 
 equivalence 
 after orbifold reduction.} 
  \label{triality}
\end{figure}

  {\bf Symplectic vs orthogonal gauge groups.} 
 Let us  now clarify how the negative dimensionality theorems impact on our models. On the level of the local symmetries in  the models with Majorana fermions \eq{LM}, \eq{LM2}, the relations \eqref{SoSp} interchange orthogonal and symplectic gauge theories.  When applied to the global $SU(\Nf)$ symmetry  the transposition of global representations accounts for the different symmetrisations of  the scalars, interchanging  $H_{(ij)}$ with  $H_{[ij]}$, and all of this accompanied by the analytic continuation of field multiplicities towards negative values
 \beq \label{minusN}
 N \mapsto -N\,,\quad \Nf \mapsto -\Nf\,.
 \eeq
Fingerprints of the negative dimensionality theorems  \eq{SoSp} can be seen on the level of the renormalisation group equations. For theories with Majorana fermions, we have confirmed at the leading orders in perturbation theory that the beta functions for the gauge, Yukawa and quartic couplings  of the theory \eq{LM} are {\it identical} to the beta functions of the theory \eq{LM2}  for any  $N$ and $\Nf$, provided we make the replacement \eq{minusN}  in the later together with  $\{g^2,y^2,u,v\} \mapsto \{-g^2,-y^2,-u,v\}$. This implies that the gauge, Yukawa, and the single and double trace quartic `t~Hooft couplings  
 \beq \label{invariant}
 \{N\, g^2,\, N\,y^2,\, \Nf\,u,\,\Nf^2\, v\}
 \eeq 
 are strictly invariant and remain positive even within the theory which has negative $N$ and $\Nf$, as they must \cite{Dyson:1952tj}. For   pure quantum gauge theories  the invariance of  $N\,g^2$   under \eq{SoSp}  is  explained  in \cite{Mkrtchian:1981bb}. Moreover, we have also confirmed that the exact same equivalence holds true for beta functions and  running couplings in theories with Dirac fermions \eq{LD} coupled to orthogonal or symplectic gauge fields. In summary, we conclude that the negative dimensionality theorems manifest themselves in the quantum theory  through the equivalence of  `t~Hooft couplings and their beta functions for any $N$ and any $\Nf$. We expect this equivalence to  hold true 
  to any order in the perturbative loop expansion. 
 
 \begin{figure}[t]
  \includegraphics[width=.65\columnwidth]{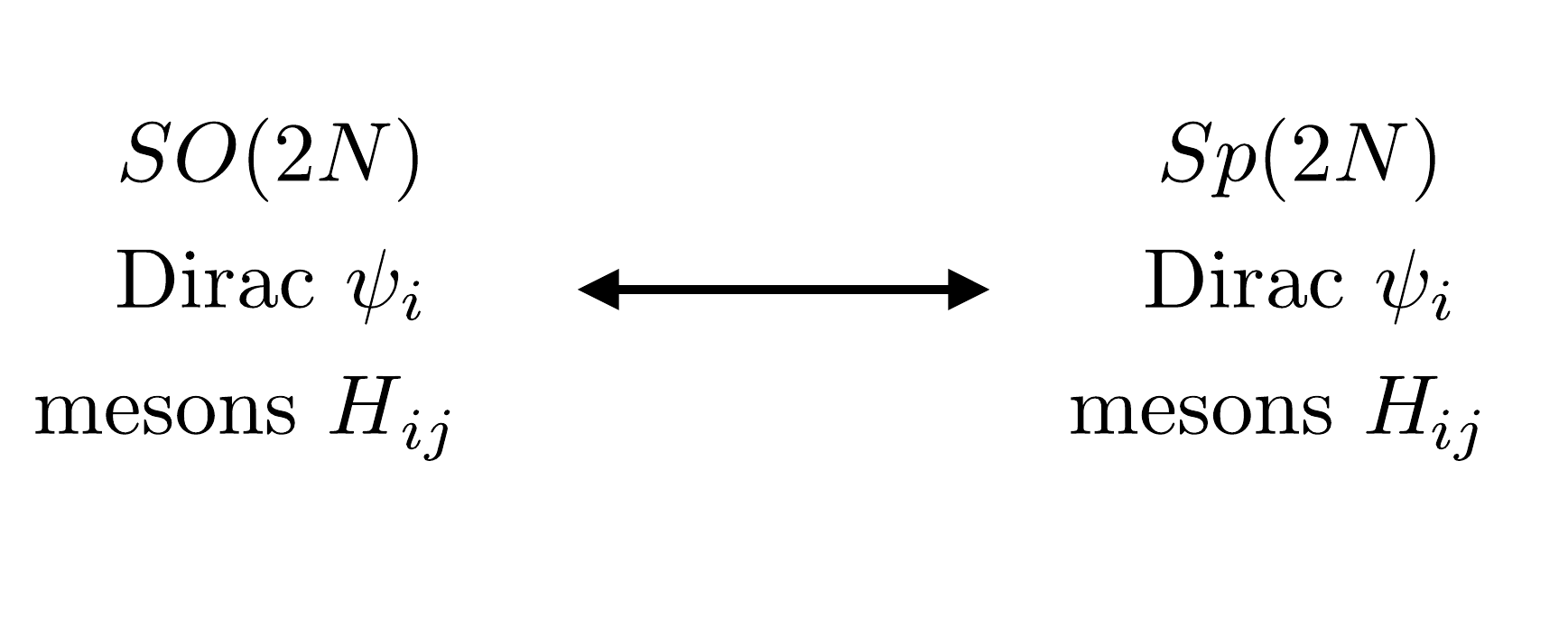}
 \vskip-.3cm
 \caption{
  Duality of  conformal fixed points, and large~$N$ equivalences of 
  matter-gauge theories with global $SU(\Nf)\times SU(\Nf)$
  flavour symmetry.}
  \label{duality1}
  \end{figure}

  \begin{table*}
\begin{center}
\begin{tabular}{ccccccccc}
\toprule
\rowcolor{LightRed}
Model
&\cell{c}{Lagrangian}
&\cell{c}{Gauge \\[-0.3em] symmetry}
&\cell{c}{Global \\[-0.3em] symmetry}
&\cell{c}{Gauge \\[-0.3em] bosons}
&\cell{c}{Fermion \\[-0.3em] type}
&\cell{c}{Weyl \\[-0.3em] components}
&Scalars
& \cell{c}{ Real scalar \\[-0.3em] components}
\\
\midrule
1
&Eq.~(\ref{LM})
& $SO(2N)$
& $SU(2 \Nf)$
& $2N^2$
&Majorana
&$2N\cdot 2 \Nf $
&$H_{(ij)}$
&$4 \Nf^2 $
\\
\rowcolor{LightGray}2
&Eq.~(\ref{LM2})
& $Sp(2N)$
& $SU(2 \Nf)$
& $2N^2$
&Majorana
&$2N\cdot 2 \Nf $
&$H_{[ij]}$
&$4 \Nf^2 $
\\ \midrule
3
&Eq.~(\ref{LD})
& $SU(N)$
& $SU(\Nf)\times SU(\Nf)$
& $N^2$
&Dirac
&$N\cdot 2\Nf $
&$H_{ij}$
&$2\Nf^2 $
\\ \midrule
\rowcolor{LightGray}4
&Eq.~(\ref{LD})
& $SO(2N)$
& $SU(\Nf)\times SU(\Nf)$
& $2N^2$
&Dirac
&$2N\cdot 2\Nf $
&$H_{ij}$
& $2 \Nf^2 $
\\
5
&Eq.~(\ref{LD}) 
& $Sp(2N)$
& $SU(\Nf)\times SU(\Nf)$
& $2N^2$
&Dirac
&$2N\cdot 2\Nf $
&$H_{ij}$
&$ 2 \Nf^2 $
\\
\bottomrule
\end{tabular}
\end{center}
 \vskip-.5cm
 \caption{\label{dofs} Gauge, fermionic, and scalar degrees of freedom in the Veneziano limit  of   models discussed in the main text.}
 \vskip-.3cm
\end{table*}

  For positive $N$ and $\Nf$, the  large~$N$ equivalence of the theories \eq{LM} and \eq{LM2}, and  of the theories \eq{LD} with either $SO$ or $Sp$ gauge symmetry, is now simple to understand,  the key point being that  the explicit dependence of beta functions  on field multiplicities arises, in the Veneziano limit, only through the parameter  $\eps$ given in \eq{eps}. Since $\eps$  is insensitive to the combined sign change  \eq{minusN}, the mapping of negative field  multiplicities in the partner theory back to positive ones  leaves all beta functions for `t~Hooft couplings \eq{invariant} invariant. The price to pay (for having positive field multiplicities on either side of the duality) is that the equivalence holds only in the large~$N$ limit. In fact,   in either  theory  the subleading corrections  start at order $1/N$ and $1/\Nf$ and enter with the same magnitude but opposite signs (once more owing to the negative dimensionality theorems) thus   breaking the duality beyond large~$N$. This pattern explains   the  equivalence of beta functions  for `t~Hooft couplings as well as   the structure of subleading corrections  found in the previous sections, and  illustrated in Figs.~\ref{triality} and~\ref{duality1}. 
 
 There are two further points worth noting with regards to   large~$N$ equivalences. First,  counting the number of gauge fields, Weyl fermions, and real scalar fields in either of these, we find that dual theories  have the exact same number of degrees of freedom  (see Model 1 vs Model 2, and Model 4 vs Model 5 in Tab.~\ref{dofs}). This  no longer holds true beyond large~$N$. Second, we also emphasize  that dual theories described here, in all cases, have the  same global symmetry but different gauge symmetry. This supports  the view that  global symmetry is a property of the system, whereas gauge symmetry is a property of the description of the system  \cite{Witten:2017hdv}.\step

 {\bf Unitary gauge groups.} 
  For the theories with Dirac fermions \eq{LD} and $SU$ gauge symmetry we confirm that  beta functions for `t~Hooft couplings are mapped onto themselves under the replacement  \eq{SoSp}, \eq{minusN}, valid for all $N$.   Moreover, subleading corrections  in the large~$N$ limit arise as inverse even  powers of field multiplicities $1/N^2$ and $1/\Nf^2$  and are insensitive to a change in sign \eq{minusN}, meaning that the theory is effectively self-dual and mapped onto itself for any $N$, $\Nf$, as it must. The result generalises to $SU$ gauge theories with  matter sectors different from \eq{LD}. 
  \step

{\bf Orbifold equivalence.}  
We now turn to the equivalence of theories between Dirac fermions coupled to unitary gauge fields, and Majorana fermions coupled  to orthogonal or symplectic gauge fields, illustrated in Fig.~\ref{triality}. Theories have different global symmetries, and those with Dirac fermions contain exactly half as many gauge, Weyl, and scalar degrees of freedom as those with Majorana fermions   in the Veneziano limit  (see Models 1 and 2 vs Model 3 in Tab.~\ref{dofs}). Still, after the identification of couplings via the map \eqref{super-universality}, all three theories have identical beta functions, phase diagrams, conformal fixed points, and  scaling exponents. This pattern  suggests that the theories are related by orbifolding.

Orbifold projections in quantum field theory  link a parent theory to a child theory with the help of  a discrete subgroup of the parent's global symmetry \cite{Bershadsky:1998cb,Schmaltz:1998bg,Erlich:1998gb,Strassler:2001fs,Armoni:2003gp,Kovtun:2004bz,Kovtun:2007py,Hanada:2011ju} (see \cite{Lucini:2012gg,Dunne:2016nmc} for reviews). "Orbifolding"   eliminates those degrees of freedom from the parent theory which are not invariant under the discrete subgroup, leading to the child theory. At the perturbative level, orbifold equivalence is based on the observation that planar diagrams of the parent and child theories coincide to all loop orders, possibly up to a rescaling of couplings, and  that  correlation functions of gauge-invariant operators  obey the same set of closed equations  \cite{Bershadsky:1998cb}. In has also been shown that the equivalence of theories holds non-perturbatively  as long as  the global symmetry used for the orbifolding is not broken spontaneously  \cite{Kovtun:2007py}.

In the setting illustrated in Fig.~\ref{triality}, the Majorana models \eq{LM} and \eq{LM2} with $SO(2N)$ or $Sp(2N)$ gauge symmetry and $SU\left(2\Nf\right)$ global symmetry  represent parent theories. Then, using a suitable $\mathbb{Z}_2$ symmetry in  the gauge and flavour groups \cite{Hanada:2011ju} leads in both cases to the  child theory  \eq{LD} with Dirac fermions, $SU(N)$ gauge symmetry,  and $SU\left(\Nf\right) \times SU\left(\Nf\right)$ global symmetry \cite{Steudtner:2019}. In the Veneziano limit, the orbifold equivalence between parent and child theories is exact, explaining the links observed in  Fig.~\ref{triality}. The factor of two which appears in the rescaling  \eqref{super-universality}  reflects that the parent theories contain twice as many gauge, Weyl, and scalar degrees of freedom as the child theory, see Tab.~\ref{dofs}.

In the literature, some orbifold/orientifold reductions  have been reported which relate supersymmetric with non-supersymmetric theories \cite{Armoni:2003gp,Dunne:2016nmc}. On the account that asymptotic safety   in supersymmetry necessitates the gauge group to be semi-simple \cite{Bond:2017suy,Martin:2000cr}, however, we do not expect to  find a supersymmetric parent for the non-supersymmetric theories with weakly coupled ultraviolet fixed points and simple gauge group studied here.\step 

{\bf Dirac vs Majorana fermions.}  Another  important observation of this study is that $SO$ and $Sp$ gauge theories with Majorana fermions  and elementary mesons can develop asymptotically safe UV fixed points while their counterparts with Dirac fermions cannot. To appreciate the  origin for this we write the leading loop contributions to the gauge and Yukawa beta functions as $\partial_t\alpha_g=\alpha^2_g(-B+C\alpha_g-D\alpha_y)$ and $\partial_t\alpha_y=\alpha_y(E\alpha_y-F\alpha_g)$ with $C,D$  and $B,E, F$ denoting  universal  two-loop  and one-loop coefficients respectively.  A necessary condition for weakly interacting UV fixed points is  given by \cite{Bond:2016dvk}
\beq \label{C'} C'\equiv C-D\,F/E<0\,.\eeq
A generic asymptotically non-free  theory ($B<0)$ has loop coefficients $C,D,E, F>0$ \cite{Bond:2016dvk}, implying  $C'\le C$.
The condition  \eq{C'} states that asymptotic safety   requires the  (Yukawa-shifted) two loop term to become negative, $C'<0$.
In all theories with exact asymptotic safety (meaning \fp2 in Tab.~\ref{FPs})  we find the universal shift
\beq \label{AS} C'/C=-\frac{38}{325}\,,
\eeq
assuming  small $0<\eps\ll 1$. Hence, the Yukawa interactions  roughly induce  a $-112\%$ correction to the two-loop gauge coefficient, which is large enough to change the sign  of $C$ and to enable asymptotic safety. 

Replacing Majorana by Dirac fermions in the theories with $SO$ or $Sp$ gauge symmetry effectively changes the scalar matter content. In fact, adjusting $N$ and $\Nf$ such that theories display the same number of gauge fields and Weyl fermions, we find that the settings with Dirac fermions only feature half as many scalar degrees of freedom (see Models 1 and 2 vs Models 4 and 5 in Tab.~\ref{dofs}). Although scalars are gauge singlets, they propagate in loops and modify  the  Yukawa loop coefficient  $E$ which  is proportional to the number of degrees of freedom   \cite{Bond:2019}.   
 Here,  the coefficient $E|_{\rm Majorana}\sim 2\Nf+2N$ reduces down to $E|_{\rm Dirac}\sim \Nf+2N$  and gives the first term in $\beta_y$ of \eq{rges:NcSO} and \eq{rges:DcSO}, respectively,  after `t~Hooft normalisation. With all other coefficients  untouched,  we find 
 \beq \label{AF} 
 C'/C=\frac{4}{125}
\eeq
instead of \eq{AS}. This corresponds  to a $-97\%$ correction of the two-loop gauge coefficient $C$, which is narrowly too small to change the overall sign of $C$. The result \eq{AF}   explains why models with Dirac fermions and $SO$ or $Sp$ gauge symmetry may display interacting {infrared}  fixed points, but cannot develop  interacting {ultraviolet} ones, much unlike their counterparts with Majoranas.\step

As a final remark, we note that the ratio  \eq{AF} also dictates the ratio of gauge couplings at the  Banks-Zaks fixed point compared to the gauge-Yukawa fixed point in asymptotically free $SO$ and $Sp$ gauge theories with Dirac fermions \eq{LD}. There,  we found that  $\alpha_g^{\rm BZ}/\alpha_g^{\rm GY}|_*=C'/C$ provided that  $0<-\eps\ll 1$ (see \fp1 and \fp3 in Tab.~\ref{FPs}). Hence, the parametric smallness of the ratio of fixed point couplings, as observed in Fig.~\ref{fig:noAS-phase}, can now be attributed to the "near-miss" of asymptotic safety due to \eq{AF}.

\section{\bf Discussion and conclusions}

As a  proof of principle,  we have established that asymptotic safety  arises in matter-gauge theories with Majorana fermions, and in theories with $SO$ and $Sp$ gauge symmetry \eq{LM}, \eq{LM2}. Together with the earlier discovery  of asymptotic safety with Dirac fermions in  $SU$ gauge theories \eq{LD}, our results clarify  that  interacting ultraviolet fixed points  can readily be realised for either type of fermions and for any of the classical gauge groups. Intriguingly though, $SU$ gauge symmetry does require fermions to be Dirac, whereas  $SO,\,Sp$ gauge symmetry does require fermions to be Majorana.

We have also put forward new classes of large $N$ equivalences between seemingly different gauge-matter theories. Equivalences between pure $SU$, $SO$ and $Sp$ gauge theories in the planar limit have been known for a long time. Here, we have explained how equivalences arise amongst theories with different local symmetries, different matter content, and, possibly, different global symmetries. Invariably,  they imply identical all-order RG flows, phase dia\-grams, and conformal critical points (Tab.~\ref{FPs}).
 Examples  (Tab.~\ref{dofs}) include a triality of  asymptotic safety (Figs.~\ref{fig:AS-phase} and \ref{triality}) or dualities amongst theories with identical infrared critical points (Figs.~\ref{fig:noAS-phase} and \ref{duality1}). 
Based on the underlying structure, many more large~$N$ equivalences arise in  gauge  theories with matter and Yukawa interactions, also offering new directions for orbifold reductions  \cite{Dunne:2016nmc,Steudtner:2019}.

Finally,  we note that our theories, at interacting fixed points,  correspond to unitary conformal field theories. This  link allows the extraction of  conformal data 	such as  scaling dimensions \cite{Litim:2014uca,Bond:2017tbw,Bond:2017lnq,Bond:2017suy} or  structure coefficients   \cite{Codello:2017hhh} directly from the  renormalisation group \cite{cardy_1996}, and in a manner complementary to the conformal bootstrap \cite{Poland:2018epd}.  It will be interesting to see whether the equivalences discovered here  (Figs.~\ref{triality} and \ref{duality1}) extend to all conformal data. This is left for future work.

\bibliographystyle{apsrev4-1}
\bibliography{PriceASbib}

\end{document}